%% file: main.tex
\documentclass[hidelinks, 12pt]{article}

\usepackage[small,bf]{caption}
\usepackage{fullpage}
\usepackage[utf8]{inputenc}
\usepackage{appendix}
\usepackage{graphicx}
\usepackage{amsmath}
\usepackage{amssymb}
\usepackage{amsthm, amsfonts}
\usepackage{subfig}
\usepackage{hyperref}
\usepackage{float}
\usepackage{tikz}
\usepackage{pgfplots}
\usepgfplotslibrary{fillbetween}

\let\phi\varphi

\newcommand{\eps}{\varepsilon}

\newcommand{\downto}{\downarrow}

\newcommand{\dint}{\mathop{\mathbf{dint}}}

\input defs.tex

\title{Improved Price Oracles: Constant Function Market Makers}
\author{Guillermo Angeris\\
{\small \texttt{angeris@stanford.edu}}
\and
Tarun Chitra\\
{\small \texttt{tarun@gauntlet.network}}}

\date{June 2020}

\begin{document}

\maketitle

\begin{abstract}
	Automated market makers, first popularized by Hanson's logarithmic market scoring rule (or LMSR) for prediction markets, have become important building blocks, called `primitives,' for decentralized finance. A particularly useful primitive is the ability to measure the price of an asset, a problem often known as the pricing oracle problem. In this paper, we focus on the analysis of a very large class of automated market makers, called constant function market makers (or CFMMs) which includes existing popular market makers such as Uniswap, Balancer, and Curve, whose yearly transaction volume totals to billions of dollars. We give sufficient conditions such that, under fairly general assumptions, agents who interact with these constant function market makers are incentivized to correctly report the price of an asset and that they can do so in a computationally efficient way. We also derive several other useful properties that were previously not known. These include lower bounds on the total value of assets held by CFMMs and lower bounds guaranteeing that no agent can, by any set of trades, drain the reserves of assets held by a given CFMM.
\end{abstract}

\section{Introduction}
As natively digital assets continue to grow, there is an increasing need for mechanisms of exchange similar to those found in traditional financial markets.
These digital assets, which include online ad impressions, cryptocurrencies, and prediction market bets, are often complex to interact with and suffer from low liquidity~\cite{othman2013practical}.
%This means that users of these assets must update their positions in these digital assets more frequently than in their traditional counterparts.
Over the last decade, a number of designs for automated market makers (AMMs) have been proposed to reduce this complexity.
AMMs encourage passive market participants with low time preference to lend their digital assets to asset pools.
The assets are then priced via a \emph{scoring rule} which maps the current pool sizes to a marginal price.
One of the most popular scoring rules is Hanson's \emph{logarithmic market scoring rule} (LMSR)~\cite{hanson2003combinatorial}.
This rule has been implemented in numerous online settings including online ad auctions~\cite{google_2017}, prediction markets~\cite{bene_2018, othman2010automated}, and instructor rating markets~\cite{chakraborty2013instructor}.

Recently, the cryptocurrency community has constructed alternative automated market makers to the LMSR (and its counterparts), known as the \emph{constant function market makers} (CFMMs).
Examples of CFMMs include Uniswap's constant product AMM~\cite{uniswap} and the constant mean AMM~\cite{balancer}.
Most applications of CFMMs have been to construct decentralized exchanges, which allow for the exchange of security-like assets without the need for a trusted third-party.
One main concern for users of these markets is whether prices on decentralized exchanges accurately follow those on centralized exchanges, which currently have more liquidity.
If the price of a decentralized exchange prices matches external prices, then such an exchange is said to be a good \emph{price oracle} that other smart contracts can query as a source of ground truth.

%Prediction markets are often used as a component of generalized oracles that aim to convey external data to a smart contract.
%We will first describe prediction markets and their connection to the generalized oracle problem, thus providing some historical context.
%Afterwards, we will discuss how CFMMs have served as decentralized exchanges with billions of dollars of annual trading volume.
%Finally, we will discuss how CFMMs, as specialized oracles, outperform generalized oracles on the task of price prediction.

\paragraph{The oracle problem.}
Of the AMM use-cases, prediction markets are the most studied and likely most controversial applications of markets implementing the LMSR.
As prediction markets have repeatedly been shut down or discontinued due to intervention by governments, large-scale stress testing of these mechanisms has been limited~\cite{phillips_2013}.
The advent of smart contract systems, such as Ethereum~\cite{wood2014ethereum} and Tezos~\cite{goodman2014tezos}, has allowed for the design and implementation of decentralized and censorship resistant versions of prediction markets.
In particular, Augur~\cite{peterson2015augur} and Gnosis~\cite{bene_2018} are smart contracts that implement LMSR-AMMs that are deployed to Ethereum's live network.
This means that anyone who owns Ethereum's native asset (ETH) can participate in these prediction markets and bet without the risk of government intervention and asset seizure (\eg, akin to `Black Friday' within the online poker community~\cite{rose2011poker}).

However, the main difficulty in using these systems in the decentralized setting is the ability to query data external to the smart contract.
For example, a prediction market betting on the future weather in Seattle would need to know the resulting weather report, once the event has happened.
%A smart contract implementing the LMSR mechanism can only achieve this if it has received external data that provides the Seattle weather reading.
Doing this in a trustless manner is difficult, as participants who have a losing bet are encouraged to try to dupe the smart contract, \ie, to manipulate the response of the contract's query.
In our previous example, if the prediction market bets on the question ``will the weather in Seattle be greater than $25^{\circ}$C?'' then a malicious participant with an active bet on ``no'' is incentivized to manipulate the query response to say that the temperature is less than $25^{\circ}$C. 
In the cryptocurrency community, the problem of providing external data to a blockchain is known as the \emph{oracle problem}, as an homage to oracles queried in theoretical computer science~\cite{koblitz2015random}.

\paragraph{Decentralized oracles.} Formally, an \emph{oracle} refers to any computational device that provides the smart contract data external to the underlying blockchain~\cite{peterson2015augur}.
There are two types of oracles in smart contract prediction markets: centralized and decentralized.
Centralized oracles involve a trusted individual or organization that provides data to the smart contract.
Examples of centralized oracles include Provable/Oraclize~\cite{mohanty2018advanced, Provable}, Wolfram Alpha~\cite{wolfram_2018}, and the MakerDAO oracle~\cite{team2017dai}. 
If these oracles are used to decide on LMSR prediction market events, they still rely on participants trusting that the centralized authority will not manipulate the data determining the final outcome of the market.
For highly valuable markets, such as the prediction of the US presidential election, it is usually untenable to trust a single individual or organization and one defers to \emph{decentralized oracles}.

Decentralized oracles are smart contracts that rely on users voting on particular prediction market outcomes.
A final outcome is chosen via a social choice function~\cite{o2014analysis}, similar to how majority or weighted majority voting is used to decide outcomes in elections.
In the case of smart contracts, the social choice function is usually significantly more complicated as the voting mechanism needs to account for adverse selection, bribery, and collusion amongst voters.
In order to reduce the likelihood of such \emph{oracle manipulation}, decentralized oracle designs often have exit games and/or complicated multiparty games that allow for certain users to challenge votes that they dispute.
Moreover, to encourage a large swath of potential users to participate in a vote, prediction market smart contracts usually provide users with a reward.
This reward is disbursed in a manner similar to how cryptocurrency rewards are given to miners and/or validators~\cite{chen2019axiomatic}.
These oracles are difficult to design as one has to balance mechanism complexity with provable defenses against collusion between prediction market participants.
Examples of decentralized oracles include Augur, Astraea~\cite{adler2018astraea}, Gnosis, and UMA~\cite{lambur_lu_cai_2019}.
We note that a stated design goal of Augur is for the prices implied by the LMSR to be used as an oracle input into other smart contracts.
For instance, if another smart contract relies on the probability of whether Seattle's temperature is greater or less than $25^{\circ}$C, then that contract simply has to subscribe to the data and pricing provided by an Augur market.
In this way, prediction market smart contracts aim to serve as the single source of off-chain data that is accessible to an arbitrarily large number of on-chain smart contracts.

\paragraph{Decentralized exchanges.}
Decentralized exchanges (DEXs) provide a method for participants to trade pairs of on-chain assets without ever needing to trust a centralized authority~\cite{0x, juliano2017dydx}, while additionally providing a means for measuring the relative price of this pair of assets. (For example, one simple but effective way in which these exchanges can provide a price would be to report the price at which the last trade was executed.)
Currently, there are roughly \$100 million of digital assets locked in DEXs with trading volume often surpassing \$10 million per day~\cite{defipulse, dex_metrics_2020}.
A design for a secure decentralized exchange for cryptocurrencies has been desired almost since the advent of Bitcoin, since centralized exchanges such as Mt.\ Gox~\cite{decker2014bitcoin}, Quadriga~\cite{shane2019crypto}, and Bitfinex~\cite{kaminska2016bitcoin} have had catastrophic losses that aggregate to billions of dollars of depositors' funds.

Many decentralized exchanges have been proposed, each with specific trading and pricing mechanisms. These range from classic order book mechanisms~\cite{0x} to other, more complicated cases~\cite{bancor}.
Yet, Uniswap~\cite{uniswap, adams_2019, angeris2019analysis}, an AMM whose pricing mechanism for comparing two assets is relatively simple in both theory and practice, has become an extraordinarily popular decentralized exchange, as measured by total trading volumes and total funds in their reserves~\cite{dex_metrics_2020}. This has led other protocols such as Celo~\cite{celo} to use the Uniswap mechanism as a price oracle.

\begin{figure}
    \centering
    \includegraphics[scale=0.9]{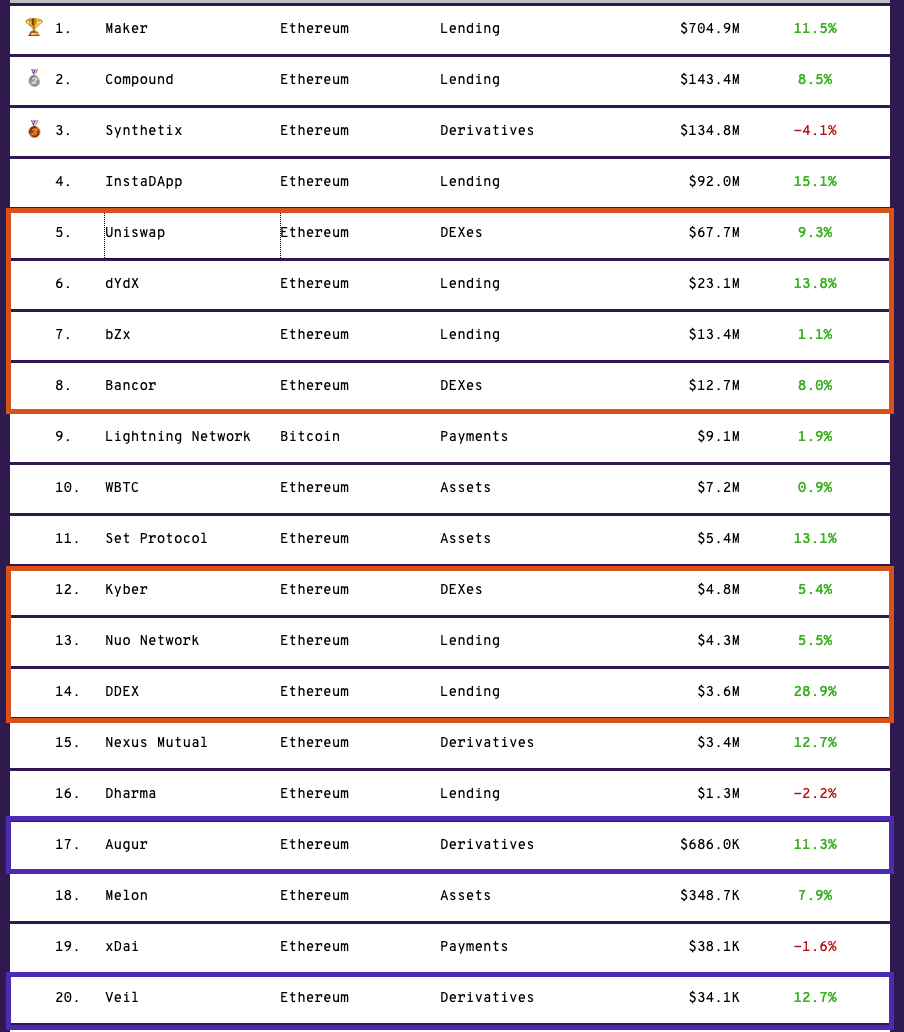}
    \caption{Market sizes of protocols utilizing constant function market makers (in red) and LMSR-based AMMs (in blue) on February 12, 2020. There are \$129.7 million of digital assets held by CFMM contracts and \$0.721 million of assets held by LMSR-based market makers. Data taken from DeFi Pulse~\cite{defipulse}.}
    \label{fig:contract_comp}
\end{figure}

\paragraph{Generalizations of Uniswap.}
The success of Uniswap, which required far fewer resources to develop than competing decentralized exchanges\footnote{Uniswap was created with a \$100,000 grant from the Ethereum Foundation~\cite{adams_2019}, whereas 0x~\cite{higgins_2017} and Bancor raised \$24 million and \$153 million in Initial Coin Offerings, respectively. On-chain data provider Dune Analytics reports that Uniswap had \$370,703,652 of volume in 2019, whereas 0x had \$228,271,968 of volume and Bancor had \$154,427,142 of volume \cite{haga, haga2}. Note that na\"ively comparing liquidity (as oppposed to volume) is more involved, as 0x currently routes orders to Uniswap and a number of aggregators exist that partition orders between different exchanges~\cite{cheng_2020}.} has led to a number of generalizations.
A natural first question is whether the Uniswap constant product formula is optimal for all types of assets.
For instance, a pair of assets which have a common mean price but different volatilities can incur large losses for the corresponding market makers and liquidity providers at specific instances in time.

An example of such assets are \emph{stablecoins}, which are digital assets whose value is (approximately) pegged to be equal to \$1.
The price of these assets naturally fluctuate around \$1 USD, but tend to stay within a bounded range of $[1-\epsilon, 1+\epsilon]$ for some $\epsilon > 0$.
The fluctuations of these assets is dictated by their natural sources of demand and can vary greatly, even though these digital assets are all meant to represent the same real world asset.
For instance, a stablecoin that is popular in Venezuela will likely have different demand characteristics than one that is popular in China.
In order to incentivize traders, the trading mechanism should instead charge lower fees when two stablecoins are near \$1 USD and higher fees when the stablecoins are farther from \$1 USD.
This approach has been implemented in Curve (previously known as StableSwap)~\cite{egorovStableSwapEfficientMechanism}, which ensures that the trading prices around \$1 USD are relatively small, but quickly becomes expensive as coins trade away from one another.

\paragraph{Multi-coin generalizations.} Another generalization of the Uniswap mechanism involves pricing multiple assets simultaneously.
Instead of providing a scoring rule that is a function of the quantities of two assets, these scoring rules take are able to price $m$ assets in terms of a set of $n$ other assets.
This allows for users to exchange portfolios of assets for other portfolios, reducing the number of transactions that the network has to handle.
On an exchange that only allows for pairwise trades, a participant would need to do $m$ trades to a num\'eraire asset (\eg, Bitcoin or USD) and then $n$ trades from the num\'eraire to the output assets.
Multi-asset generalizations of Uniswap, such as Balancer~\cite{balancer}, would execute such a trade atomically, reducing fees and price slippage.
The choice of scoring rule affects how easy it is for arbitrageurs to keep the portfolio prices synchronized with the prices of the underlying components.

While this mechanism might seem arbitrary, there are a number of examples of similar assets from traditional finance that involve trading baskets of good for other baskets of goods.
For instance, an Exchange Traded Fund (ETF) is a single equity instrument $S_E$ that represents a weighted set of shares $S_1, \ldots, S_n$.
There are currently over \$5 trillion of assets locked in ETFs~\cite{statista}.
Many ETFs represent a share of $S_E$ by a weighted linear combination of the shares, $S_E = \sum_{i=1}^n w_i S_i$ for some positive integer weights $w_i$.
If any of the prices of shares $S_i$ change, leaving $S_E$ mispriced, an arbitrageur can perform creation-redemption arbitrage~\cite{gastineau2001introduction}.
This arbitrage works due to two steps:
\begin{itemize}
    \item Creation: A market participant can create a single share $S_E$ by providing the ETF underwriter with $w_i$ shares of $S_i$ for all $i$
    \item Redemption: A market participant can redeem a single share $S_E$ by giving the ETF underwriter $S_E$ and receiving $\{w_i S_i : i \in [n]\}$. 
\end{itemize}
If the price of $S_E$ is higher than the weighted sum of the prices of $S_i$, then an arbitrageur can buy the basket $\{w_i S_i : i \in [n]\}$ for less than $\text{price}(S_E)$, create a share of $S_E$ and sell it for a profit of $\text{price}(S_E) - \sum_{i=1}^n w_i \cdot \text{price}(S_i)$.
Similarly, the arbitrageur can use a redemption to arbitrage a low $S_E$ price.
Multi-asset generalizations of Uniswap allow for precisely this type of arbitrage to occur without the need for a trusted intermediary (\eg, the ETF underwriter).

%\begin{figure}
%    \centering
%    \includegraphics[width=0.45\textwidth]{curve}
%    \caption{A comparison of StableSwap/Curve and Uniswap. Figure taken from~\cite{egorovStableSwapEfficientMechanism}.}
%    \label{fig:curve}
%\end{figure}

\paragraph{Summary.}
In this paper, we show that many of the generalizations presented are special cases of a large family of \emph{constant function market makers}, or CFMMs, which all satisfy relatively similar and useful theoretical properties, under mild conditions. In particular,
we give a complete and very general framework for analyzing CFMMs that includes all of the examples of CFMMs currently used
in practice. We also provide sufficient conditions for these CFMMs to be well-behaved, in the sense that
(a) agents are incentivized to have the CFMM correctly report the price of its assets, when compared to a reference market, and
(b) agents can never drain the assets that a CFMM contains by only trading with the given CFMM. Additionally, we give some simple derivations
of a few properties of interest, including the total asset value that a given CFMM holds, as a function of the external market prices. Finally, we provide some simple extensions of the current analysis and possible future directions for a more general analysis of CFMMs.

\section{Constant function market makers}
\label{sec:cfmm-multi}
In this section, we will discuss the basic definitions of constant function market makers and how Uniswap, among other CFMMs, fit
directly into the provided framework.

\paragraph{Definition of a CFMM.} A \emph{constant function market maker}, or CFMM, is a type of automated market maker defined by its \emph{trading function},  $\phi : \reals^n_+ \times \reals^n_{+} \times \reals^{n}_{+} \to \reals$, and its \emph{reserves}, $R \in \reals^n_+$. Here, $R_i$ specifies how much of coin $i$ the CFMM is allowed to use or interact with, while the trading function specifies what constitutes a valid trade.

%We will call the function specifying the CFMM behavior a trading function rather than a scoring rule (as is the case with the LMSR) to prevent possible confusion and to emphasize that, while superficially similar, these are rather distinct concepts. For example, a trading function's value need not represent any useful quantity nor a probability distribution over such quantities, unlike scoring rules, and it is also not required to satisfy any conditions. (Our analysis will only assume a convexity of sorts, as given in~\S\ref{sec:convexity}, which, while considerably general, does not technically include all possible CFMMs.) We additionally discuss why CFMMs and scoring rules in the traditional sense cannot be generally compared under the rather general assumptions we use in this paper in~\S\ref{sec:lmsr-comparison}.

In all of the applications we will see, a decentralized smart contract will implement a given CFMM. In this scenario, the CFMM will have some balance of various tokens available to it (with the $i$th entry of the reserves vector $R$ specifying the amount of token $i$ that is available), and agents can interact with this CFMM in various ways.

\paragraph{Agents and actions.} In one case, a \emph{liquidity provider} provides some number of funds to the reserves of the contract. In return, the liquidity provider is given a form of IOU (usually in the form of other tokens) which can be later redeemed for some fixed percentage of the reserve amounts.

In contrast, a \emph{trader} attempts to exchange some given amount of coins in reserves. For example, a trader may simply want to trade
some amount of coin $i$ for coin $j$, in which case the trader would deposit some amount of coin $i$ into the reserves of the contract and withdraw some amount of coin $j$ from these reserves.

\paragraph{The trading function.} More generally, a trader may wish to trade any number of coins for any other coins in potentially very complicated ways. To specify this, we will write the \emph{output trade}, $\Lambda \in \reals^{n}_+$, as a vector with $n$ nonnegative entries. (We will often simply call this the \emph{output}.) The output's $i$th entry, $\Lambda_{i}$, specifies how much of coin $i$ the trader wishes to receive from the CFMM. Additionally, we will define the \emph{input} $\Delta \in \reals_+^n$, which is a vector whose $i$th entry, $\Delta_i$ specifies how much of coin $i$ the trader has given the contract.  We will call the resulting tuple, $(\Delta, \Lambda) \in \reals^n_+\times \reals^n_+$, a \emph{trade}.

The trading function $\phi$ then specifies exactly when a trade is considered valid and therefore executed. In particular, the smart contract with reserves $R$ only accepts the trade given by $(\Delta, \Lambda)$ whenever $(\Delta, \Lambda)$ satisfies
\begin{equation}\label{eq:feasible-trade}
\phi(R, \Delta, \Lambda) = \phi(R, 0, 0).
\end{equation}
In other words, the trade is accepted only if the trading function is kept constant.
The contract then places the coins $\Delta$ provided by the trader into its reserves while paying out $\Lambda$ to the trader. This results in the reserves being updated in the following way:
\[
R \gets R + \Delta - \Lambda,
\]
after a valid trade is executed.

\paragraph{Examples.} Throughout the remainder of this paper, we will use constant product markets (Uniswap) and constant mean markets (Balancer) as mathematically simple but very practical and useful instances of CFMMs and as the canonical examples for the definitions and derivations provided. 

In the case of constant product markets with percentage fee $(1-\gamma)$ (see, \eg,~\cite[\S2]{angeris2019analysis}), we have $n=2$ and the trading function is:
\begin{equation}\label{eq:uniswap-cfmm}
\phi(R, \Delta, \Lambda) = (R_1 + \gamma\Delta_1 - \Lambda_1)(R_2 + \gamma\Delta_2 - \Lambda_2).
\end{equation}
Of course, no rational trader will ever opt to make both $\Delta_1 \ne 0$ and $\Lambda_{1} \ne 0$ in the case of nonzero fees; \ie, a rational trader will
never trade a specific coin for a smaller amount of the same coin. The same is true for the pair $\Delta_2$ and $\Lambda_2$. This results in the more recognizable form:
\[
\phi(R, \Delta, \Lambda) = \begin{cases}
(R_1 + \gamma\Delta_1)(R_2 - \Lambda_2) & \Delta_2 = \Lambda_1 = 0\\
(R_1 - \Lambda_1)(R_2 + \gamma\Delta_2) & \Delta_1 = \Lambda_2 = 0\\
\phi(R, 0, 0) + \eps & \text{otherwise}.
\end{cases}
\]
(One can make a similar argument in the fee-less case with $\gamma = 1$.) Here, $\eps \ne 0$ can be any nonzero value,
as it should simply prevent any trade which has both $\Lambda_1$ and $\Delta_1$ nonzero or $\Lambda_2$ and $\Delta_2$ nonzero, such that~\eqref{eq:feasible-trade} cannot be satisfied in either of these cases. Throughout the rest of this paper, we will work with the form given by~\eqref{eq:uniswap-cfmm}, since the construction is easier to handle mathematically, but both forms are easily seen to be equivalent in the sense described in~\S\ref{sec:trading-set}
%Since we are separating the input coin from
%the output coin and allowing two-way trades, trading function~\eqref{eq:uniswap-cfmm} differs slightly from the usual treatment of Uniswap, but the construction is otherwise identical.

The CFMM for constant mean markets, originally proposed by Balancer~\cite{balancer} for $n$ coins is (\cf,~\cite[\S3]{angeris2019analysis}):
\begin{equation}\label{eq:balancer-cfmm}
\phi(R, \Delta, \Lambda) = \prod_{i=1}^n (R_i + \gamma\Delta_i - \Lambda_i)^{w_i}.
\end{equation}
Here, $0 < \gamma \le 1$ and the weights $w \in \reals^n_+$ are nonnegative and satisfy $\ones^Tw = 1$. As before, no rational trader will have $\Delta_i \ne 0$ and $\Lambda_i \ne 0$ if $\gamma < 1$, and we can always take
at least one of these to be zero for any trade, even when $\gamma = 1$.

\subsection{The trading set and implications}\label{sec:trading-set}
There are potentially many trading functions $\phi$ which are all `equivalent' in some sense or another. One example is to compare the
functions $\phi$ and $-\phi$. Clearly, any trade which is feasible in one CFMM is feasible in the other, yet the functions are not equal except when both take on the value 0. 

To fix this issue, we introduce the \emph{trading set} $T(R) \subseteq \reals_+^n\times \reals^{n}_+$ at reserves $R$, defined as
\begin{equation}\label{eq:trading-set}
T(R) = \{(\Delta, \Lambda) \mid \phi(R, \Delta', \Lambda') = 0 ~\, \text{for some} ~\, \Delta' \le \Delta,~ \Lambda' \ge \Lambda\},
\end{equation}
where the inequalities are all taken elementwise. Another way of stating this definition is the following: the trade $(\Delta, \Lambda) \in T(R)$ is in the trading set only when there exists a (potentially) cheaper alternative $\Delta' \le \Delta$, or one with (potentially) higher payoff, $\Lambda' \ge \Lambda$.

The general idea is that no rational agent will ever pick a trade $(\Delta, \Lambda) \in T(R)$ whenever there exists a better trade $(\Delta', \Lambda')$ satisfying $\Delta' \le \Delta$ and $\Lambda' \ge \Lambda$ with
at least one of the elementwise inequalities holding strictly.
%In this case we will say the trade $(\Delta', \Lambda')$ \emph{dominates} $(\Delta, \Lambda)$.
We also note that this definition of a trading set is similar, though not quite equivalent, to the idea of an epigraph in optimization theory~\cite[\S3.1.7]{cvxbook}.

The trading set contains all of the important information provided by the trading function $\phi$, for a given CFMM. Because of this, we will often opt to work with the trading set instead of the function $\phi$ directly, as working with the set is generally mathematically simpler and more elegant. In many (but not all) cases, the statements for the trading set can easily be translated to statements over the function $\phi$. When this is the case, we will write both statements directly.

%\paragraph{Dominated trade set.} We will call the set of trades in $T(R)$ that are dominated by at least one trade the \emph{dominated trade set} of $T(R)$. For a given trade set $T(R)$, this is defined:
%\[
%\dint T(R) = \{(\Delta, \Lambda) \in T(R) \mid \text{there exists} ~  (\Delta', \Lambda') \in T(R) ~\text{dominating} ~ (\Delta, \Lambda) \}.
%\]
%The dominated trade set is exactly the set of trades that are suboptimal when every coin $i$ has nonzero value and therefore no rational agent would choose such a trade.
%
%We also note that, in many cases, the dominated trade set is exactly the topological interior of the trade set (and hence the suggestive operator name $\dint$ as the \emph{d}ominated \emph{int}erior), but this need not be true in general. For instance, consider the trade set given by $\Lambda, \Delta \in \reals^2_+$ satisfying $\Lambda_1 = \Delta_1$ and $\Delta_2 \ge \Lambda_2$. This set has empty interior, but nonempty dominated trade set $\Lambda_1 = \Lambda_2$ and $\Delta_2 > \Lambda_2$. In general, it is always true that the interior of the set satisfies $\intr T(R) \subseteq \dint T(R)$. (This inclusion also holds true for the relative affine interior~\cite[\S2.1.3]{cvxbook}.) This additionally implies that showing that a point is in the interior of $T(R)$ suffices to show that it is in the dominated interior, a useful sufficient condition in many cases.

\paragraph{Feasibility and equivalence.} We will say some trade $(\Delta, \Lambda)$ is \emph{feasible} if $(\Delta, \Lambda) \in T(R)$ and define two CFMMs to be \emph{equivalent} if their trading sets are equal (\ie, if the same trades are feasible in both). The main idea is that the definition of $T$ is essentially unique with respect to the actions that any rational agent would take. More specifically, if two trading sets are not equal $T(R) \ne T'(R)$ for some reserves $R$, say, then there exists a trade in one set which is not feasible in the other; in other words, at least one trade $(\Delta, \Lambda) \in T(R)$ is not in $T'(R)$ and there exists no trade dominating $(\Delta, \Lambda)$ in either set.

An interesting and surprising consequence of this definition is that, given any trading function $\phi$, we can construct an equivalent CFMM with trading function $\phi'$ such that $\phi'(R, \cdot, \cdot)$ is monotonically nonincreasing in its first argument and nondecreasing in its second. We show this and discuss some additional implications in~\S\ref{app:monotonicity}.

\paragraph{Equivalence in practice.} The trading set for a constant mean market~\eqref{eq:balancer-cfmm} can easily be written as the set of all nonnegative $(\Delta, \Lambda)$ that satisfy
\[
\prod_{i=1}^n (R_i + \gamma\Delta_i - \Lambda_i)^{w_i} \ge \prod_{i=1}^n R_i^{w_i}.
\]
If we were to take $n=2$ and $w_1 = w_2 = 1/2$, then the trading set is the set of all nonnegative $(\Delta, \Lambda)$ that satisfy
\[
(R_1 + \gamma\Delta_1 - \Lambda_1)^{1/2}(R_2 + \gamma\Delta_2 - \Lambda_2)^{1/2} \ge R_1^{1/2}R_2^{1/2}.
\]
Squaring both sides of the inequality, as both quantities are nonnegative, gives exactly the equation for the trading set for
the constant product market~\eqref{eq:uniswap-cfmm} for any possible reserve vector $R$ and fee $\gamma \le 1$. This implies that the trading sets are equal, which, in turn, implies the CFMMs are equivalent, as expected.

In fact, this case is the special case where the two trading functions are related by a strictly monotonic transformation, so the two trading sets are obviously equivalent. (More generally, the CFMMs with trading functions $\phi$ and $f\circ \phi$ are equivalent for any invertible transformation $f:\reals\to\reals$.) An enlightening exercise is to write two different, but equivalent, trading functions for constant product and constant mean markets for which there exists no invertible transformation mapping one to the other.

\subsection{Convexity of the trading set}\label{sec:convexity}
We will make only one assumption regarding the trading set $T(R)$: for each possible reserve vector $R$, the set $T(R)$ is a closed convex set.

\paragraph{Discussion.} This assumption of convexity has two important consequences. First, the geometry of convex sets is an extremely well-studied and developed field, and
many basic results from this field suffice for the mathematical purposes of this paper, while remaining considerably general. The second case is that, when the set $T(R)$ can be written down in a compact way (\eg, as the intersection of a polynomial number of well-known convex cones), then we can usually solve convex optimization problems over this set in a computationally efficient way~\cite[\S1]{cvxbook},
even when closed-form solutions aren't guaranteed to exist. As we will show later in~\S\ref{sec:arbitrage}, this implies that agents can easily maximize their payoff by computing an appropriate solution to an optimization problem over the trading set.

%\paragraph{Sufficient conditions.} A simple sufficient condition is to note that, if the trading function is of the form
%\[
%\phi(R, \Delta, \Lambda) = \psi(R, A(\Delta, -\Lambda)),
%\]
%for some function $\psi: \reals^n\times\reals^m \to \reals$, where $A \in \reals^{m \times 2n}_+$ is a nonnegative matrix, then it suffices to prove that the set
%\begin{equation}\label{eq:set-convex}
%S(R) = \{x \in \reals^m_+ \mid \psi(R, y) = \psi(R, 0) ~\, \text{for some} ~\, y \le x\},
%\end{equation}
%is convex. This follows from the fact that $A(\Delta, -\Lambda)$ is a nondecreasing function of $\Delta$ and a nonincreasing function of $\Lambda$ as $A$ is nonnegative, along with the fact that the trading set $T(R)$ is the inverse image of $S$ under an affine transformation, which preserves convexity. This condition often serves to simplify the analysis of most of the trading functions used in practice.

\paragraph{Convexity in practice.} So far, to the knowledge of the authors at the time of writing, there is no CFMM used in practice whose trading set is nonconvex. For example, constant mean market CFMMs have convex trading sets since the weighted geometric mean function is a concave function, composition with an affine function preserves convexity, and the superlevel sets of a concave function are convex (\cf,~\cite[\S3.1]{cvxbook} for elementary proofs). This implies that the trading set
\[
T(R) = {\textstyle \{(\Delta, \Lambda) \in \reals^n_+\times \reals^n_+ \mid \prod_{i=1}^n(R_i + \Delta_i - \Lambda_i)^{w_i} \ge \prod_{i=1}^n R_i^{w_i}\}},
\]
is therefore a convex set, for any valid weights $w \ge 0$ with $\ones^T w = 1$ and reserves $R$. (This also immediately shows that constant product markets have convex trading sets, as a special case.)

For a more complicated example, we will consider the trading function for Curve~\cite{egorovStableSwapEfficientMechanism}:
\begin{equation}\label{eq:curve-cfmm}
\phi(R, \Delta, \Lambda) = \alpha \ones^T(R + \gamma \Delta - \Lambda) - \beta \prod_{i=1}^n(R_i + \gamma\Delta_i - \Lambda_i)^{-1}, 
\end{equation}
where $\alpha, \beta \ge 0$ are tunable parameters, $(1-\gamma)$ is the percentage fee, and, as usual, $(\Delta, \Lambda) \in \reals_+^n\times \reals_+^n$. We can show that the trading set is convex directly, since the function $x \mapsto -(\prod_{i=1}^n x_i)^{-1}$ is a concave function that is increasing in each of its arguments. (A simple proof follows from the fact that $x \mapsto \log\prod_{i=1}^n x_i^{-1}$ is convex as it is the sum of negative logarithms, and a function is convex if it is log-convex~\cite[\S3.5.1]{cvxbook}.) Therefore, the set of all $(\Delta, \Lambda)$ satisfying
\[
\alpha \ones^T(R + \gamma \Delta - \Lambda) - \beta \prod_{i=1}^n(R_i + \gamma\Delta_i - \Lambda_i)^{-1} \ge \alpha \ones^TR-\beta\prod_{i=1}^n R_i^{-1},
\]
is, again, a convex set as it is the superlevel set of the sum of an affine and a concave function.

\subsection{Path deficiency and path independence}
There are many special CFMMs which, when appropriately generalized, yield a very natural family of trading functions and sets worth studying. One particularly useful property that makes the analysis of some CFMMs simpler is path independence. While practical CFMMs usually don't satisfy path independence (often owing to their fee structure), path independent CFMMs often serve as a good starting point for reasoning about possible CFMMs and many of the special cases we study will be path independent.

We note that, unlike in the classical automated market maker literature (where path independence is almost a requirement), almost no CFMMs in practice are path independent. Much of the theory presented here also does not require this property to hold except in specific cases, and we will make explicit note of this when those results are presented. Because of this we will present a slight generalization of path independence, called path deficiency, which retains many of the useful properties but is satisfied by all known CFMMs used in practice.

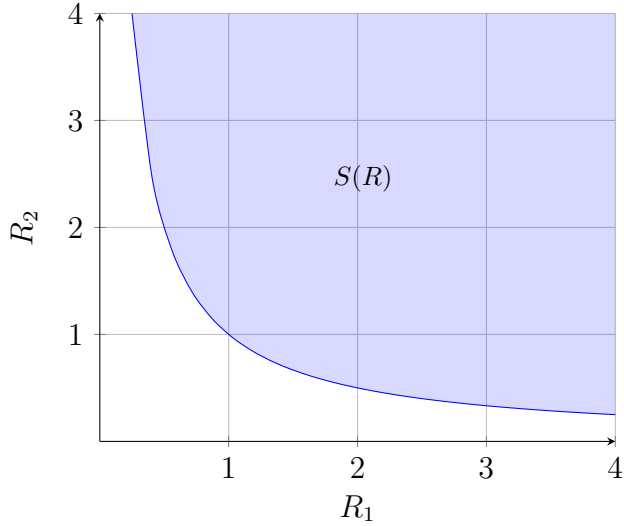
\begin{figure}
\centering
\begin{tikzpicture}
\begin{axis}[
axis lines=center, 
grid=major,
ymin=0, 
no marks,
xmin=0,
xmax=4,
xlabel=$R_1$,
ylabel=$R_2$,
x label style={at={(axis description cs:0.5,-0.1)},anchor=north},
y label style={at={(axis description cs:-0.1,.5)},rotate=90,anchor=south},
]
\addplot+[smooth,blue,name path=A,domain=0.25:4] {1/x};
\addplot+[draw=none,name path=B] {4};
\addplot+[blue,opacity=0.15] fill between[of=A and B,soft clip={domain=0.25:4}];
\end{axis}
\tikzstyle{fontbf} = [thick, font=\fontsize{10}{0}\selectfont]
\draw (3.5, 3.5) node[fontbf] {$S(R)$};
\end{tikzpicture}
\caption{Example reachable set.}
\label{fig:reachable-set}
\end{figure}

\paragraph{The reachable reserve set.} The definitions of path deficiency and path independence can be phrased in simple ways in terms of the \emph{reachable reserve set}, defined, for fixed reserves $R \in \reals_+^n$, as
\begin{equation}\label{eq:reachable-reserves}
S(R) = \{R + \Delta - \Lambda \mid (\Delta, \Lambda) \in T(R)\}.
\end{equation}
Equivalently, the reachable reserve set, $S(R)$, is the set of reserves which can be \emph{reached} from the current CFMM reserves by
performing a single feasible trade. Note that the set $S(R)$ is also convex as it is the image of $T(R)$ under an affine transformation.
The reachable reserve set is shown in figure~\ref{fig:reachable-set} for the case of constant product markets with reserves satisfying $R_1R_2 = 1$ and no fees.

\subsubsection{Path deficiency}\label{sec:path-deficiency}
We will say that a CFMM is \emph{path deficient} if, for given reserves $R$, any reachable reserve
$R' \in S(R)$ has a reachable set satisfying $S(R') \subseteq S(R)$. In other words, performing any feasible trade can only
make the reachable reserve set no larger than its current one. 

This has a very simple implication: any path deficient CFMM initialized with reserves $R^0$ will always satisfy
\[
\ones^TR \ge \inf_{R' \in S(R^0)} \ones^TR',
\]
where $R$ are the reserves after any number of feasible trades have been executed. This follows from the fact that $S(R) \subseteq S(R^0)$, which generalizes the known lower bounds for the total amount of coin in reserves for, \eg, constant product and constant mean market makers~\cite[\S2.3]{angeris2019analysis}. In other words, path deficiency guarantees that the reserves are always bounded from below after any set of trades has been performed.

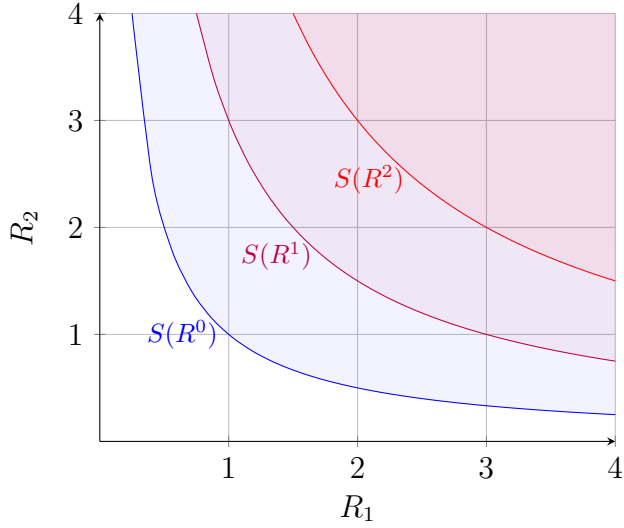
\begin{figure}
\centering
\begin{tikzpicture}
\begin{axis}[
axis lines=center, 
grid=major,
ymin=0, 
no marks,
xmin=0,
xmax=4,
xlabel=$R_1$,
ylabel=$R_2$,
x label style={at={(axis description cs:0.5,-0.1)},anchor=north},
y label style={at={(axis description cs:-0.1,.5)},rotate=90,anchor=south},
]
\addplot+[draw=none,name path=B] {4};
\addplot+[smooth,blue,name path=R0,domain=0.25:4] {1/x} node[left,pos=0.5, thick,font=\fontsize{10}{0}\selectfont] {$S(R^0)$};
\addplot+[blue,opacity=0.05] fill between[of=R0 and B,soft clip={domain=0.25:4}];
\addplot+[smooth,purple,name path=R1,domain=0.75:4] {3/x} node[left,pos=0.5, thick,font=\fontsize{10}{0}\selectfont] {$S(R^1)$};
\addplot+[purple,opacity=0.05] fill between[of=R1 and B,soft clip={domain=0.25:4}];
\addplot+[solid,smooth,red,name path=R2, domain=1.5:4] {6/x} node[left,pos=0.5, thick,font=\fontsize{10}{0}\selectfont] {$S(R^2)$};
\addplot+[red,opacity=0.05] fill between[of=R2 and B,soft clip={domain=0.25:4}];
\end{axis}
\end{tikzpicture}
\caption{Example reachable sets of a (strictly) path-deficient CFMM after two trades.}
\label{fig:path-deficient}
\end{figure}

\paragraph{Strict path deficiency.} Additionally, we will say that a CFMM is \emph{strictly path deficient} if
$R'\in S(R)$ implies that $S(R') \subseteq \dint S(R)$, where $\dint S(R)$ is the dominated interior of $S(R)$, defined as
\[
\dint S(R) = \{R' \in S(R) \mid R' ~\text{dominates some element in} ~ S(R)\}.
\]
For two nonnegative vectors, $x, y \in \reals_+^n$, we say $x$ \emph{dominates} $y$ if $x \ge y$ and at least one of the inequalities holds strictly. Clearly, a strictly path deficient CFMM is path deficient, but not vice versa. Additionally, note that $\relint S(R) \subseteq \dint S(R)$, where $\relint S(R)$ is the relative interior~\cite[\S2.1.3]{cvxbook} of $S(R)$. See, \eg, figure~\ref{fig:path-deficient} for an example.

The main idea of strict path deficiency is to note that any agent will always prefer to make a trade such that the new CFMM reserves satisfy $R' \in S(R) \setminus \dint S(R)$ to a trade which changes the CFMM reserves to $R' \in \dint S(R)$, since, by definition, $R' \in \dint S(R)$ implies that the agent either received strictly less output or had to add in strictly more input than at least one possible feasible trade in $S(R)$.

\paragraph{Sufficient condition.} A simple but very useful sufficient condition for strict path deficiency is that a CFMM is path
deficient if we can write its trading set in the following form:
\[
T(R) = \{(\Delta, \Lambda) \in \reals_+^n\times \reals_+^n \mid \psi(R + \gamma\Delta - \Lambda) \ge \psi(R)\},
\]
where  $0 \le \gamma < 1$ and $\psi$ is a quasiconcave function that is strictly increasing in its arguments.
(For more information on quasiconcave functions, see, \eg,~\cite{agrawalDisciplinedQuasiconvexProgramming2019}.)
A quick sketch of the proof is to note that $R + \Delta - \Lambda$ dominates $R + \gamma\Delta - \Lambda$
in at least one entry if $\Delta \ne 0$. Because the function is strictly increasing, then
\[
\psi(R + \Delta - \Lambda) > \psi(R + \gamma\Delta - \Lambda) \ge \psi(R),
\]
which gives the result after using the definition of the reachable reserve set~\eqref{eq:reachable-reserves} and using the fact that
$\psi(R') > \psi(R)$ and $R' \ge R$ implies that $R'$ dominates $R$.

This provides a much simpler proof of path deficiency than the one given in appendix~D of~\cite{angeris2019analysis} for constant
product markets, and this sufficient condition covers all of the example trading functions given by equations~\eqref{eq:balancer-cfmm} and~\eqref{eq:curve-cfmm}, whenever $0 \le \gamma < 1$.

%
%A proof of this is nearly immediate from the definition. Consider any $R'' \in S(R')$, then since $R'$ is, by definition reachable from $R$, and $R''$ is itself reachable from $R'$, then there exist two sequentially feasible trades $(\Delta, \Lambda) \in T(R)$ and $(\Delta', \Lambda') \in T(R')$ such that $R'' = R' + \Delta' - \Lambda' = R + \Delta - \Lambda + \Delta' - \Lambda'$. By definition, the aggregate trade is also feasible, $(\Delta + \Delta', \Lambda + \Lambda') \in T(R)$, so $R''$ is also reachable from $R$, which implies $S(R') \subseteq S(R)$.

\subsubsection{Path independence}
In a manner analogous to the classical market making literature, we will say a CFMM is \emph{path independent} if it is path deficient and its reachable reserve set does not change after any feasible trade that is not dominated by another. In other words, a CFMM with reserves $R$ is path independent whenever it is path deficient and $R' \in S(R) \setminus \dint S(R)$ implies $S(R') = S(R)$. By definition, any path independent CFMM is path deficient, but not strictly path deficient.

The main reason why path independence simplifies many of the derivations provided is that, in general, all of its properties can be written only as a function of the reserves the CFMM has at any one point in time, since no rational agent will ever make a trade such that the resulting reserves lie in the dominated interior, $\dint(R)$, as there would exist a feasible trade that would have been strictly better for the agent. We will see in the following sections that this yields a simple approach to finding the marginal price of given assets, among other cases.

\paragraph{Sufficient condition.} A simple sufficient condition for path independence is that the set $T(R)$ can be written as
\begin{equation}\label{eq:sufficient-condition}
T(R) = \{(\Delta, \Lambda) \mid \psi(R + \Delta - \Lambda) \ge \psi(R)\},
\end{equation}
where $\psi$ is a nondecreasing quasiconcave function. Since a concave function is quasiconcave, this condition is enough to show that
all of the example trading functions, given by equations~\eqref{eq:balancer-cfmm} and~\eqref{eq:curve-cfmm}, are path independent whenever they have no fees ($\gamma = 1$).

\subsection{Optimal arbitrage and the marginal price}\label{sec:arbitrage}
First, we will look at the no-arbitrage conditions for a given CFMM. We will then look at how these conditions imply a specific marginal
price for fixed reserves and then explore some important special cases for which the marginal prices are easy to write down.

\paragraph{Optimal arbitrage.} In the arbitrage problem, we have a reference market of coins $i=1, \dots, n$, where
each can be sold or purchased at some fixed price given by $c_i > 0$ for each coin $i$. In this problem, an agent
(usually called an arbitrageur) is allowed to borrow $\Delta_i$ of coin $i$ for $i=1, \dots, n$ and use it to trade with both the reference market and the CFMM,
with the only condition that the loan is repaid at the end of the trade. The \emph{optimal arbitrage problem} then asks what is the
optimal arbitrage trade; \ie, what is the optimal value of the problem
\begin{equation}\label{eq:optimal-arbitrage-trade-set}
\begin{aligned}
	& \text{maximize} && c^T(\Lambda - \Delta)\\
	& \text{subject to} && (\Delta, \Lambda) \in T(R),
\end{aligned}
\end{equation}
with variables $\Delta, \Lambda \in \reals^n$, where $R$ specifies the current reserves of the CFMM at the time the trade is to be performed. An equivalent formulation in terms of the reachable set is
\begin{equation}\label{eq:optimal-arbitrage}
\begin{aligned}
	& \text{maximize} && c^T(R - R')\\
	& \text{subject to} && R' \in S(R),
\end{aligned}
\end{equation}
with variable $R' \in \reals^n$. The equivalence follows by noting that we can write $R' = R + \Delta - \Lambda$ and there exists such a $(\Delta, \Lambda) \in T(R)$ if, and only if, $R' \in S(R)$, by definition. Additionally, note that any optimal point for~\eqref{eq:optimal-arbitrage} will always have $R' \in S(R) \setminus \dint S(R)$.

Because both problems are convex problems, it is almost always the case that they can be efficiently solved whenever the set $T(R)$ can be compactly expressed in terms of well-known convex sets. Additionally, the optimality conditions
of problem~\eqref{eq:optimal-arbitrage} imply that there exists a supporting hyperplane for the set
$S(R)$ with slope collinear to $c$ at the optimal point $R'^\star$. (For more details, see Corollary~11.5.2 and Theorem~18.8 of~\cite{rockafellar1970convex}.)

\paragraph{Path deficiency.} In general, it may be possible that solving problem~\eqref{eq:optimal-arbitrage} and executing the optimal trade provided is actually not the best possible strategy for a specific arbitrageur. For example, it may be the case that an arbitrageur could somehow have a higher-payoff strategy by breaking up the trade into many smaller trades and performing some complicated trading procedure that results in a better payoff.

In this case, the idea of path deficiency is extraordinarily useful: if a CFMM is path deficient, then there is no strategy by which
an arbitrageur could have higher payoff than simply solving problem~\eqref{eq:optimal-arbitrage-trade-set} or~\eqref{eq:optimal-arbitrage} and executing the resulting trade. Additionally, if the CFMM is strictly path deficient, then the arbitrageur only does
worse by attempting to subdivide the resulting trades.

To show this, we will consider any strategy $\{R^i\}_{i=0}^m$ such that the $i$th action taken by the arbitrageur changes the reserves from $R^i$ to $R^{i+1}$ and is feasible, \ie, $R^{i+1} \in S(R^i)$. Now, by definition, the price paid by the agent during the $i$th action is equal to $c^T(R^i - R^{i+1})$. Summing over the prices implies that the total payoff of this strategy is equal to $c^T(R^0 - R^m)$. Since the CFMM is path deficient, the reachable reserve sets satisfy $S(R^{m-1}) \subseteq \dots \subseteq S(R^0)$, so $R^m \in S(R^0)$, which implies that this strategy cannot have higher payoff than the optimal value of~\eqref{eq:optimal-arbitrage}, since the strategy taking $R^0$ to $R^m$ in one step is feasible for~\eqref{eq:optimal-arbitrage} and has the same payoff.

The case where the CFMM is strictly path deficient is rather similar. If the number of moves taken satisfies $m > 1$, then clearly $R^m \in \dint S(R^0) \subseteq S(R^0)$. But, by definition, $R^m$ is in the dominated interior only when there exists a trade $R' \in S(R^0)$ which dominates $R^m$. Because $R'$ is feasible for~\eqref{eq:optimal-arbitrage}, then the optimal value of problem~\eqref{eq:optimal-arbitrage} must be strictly larger than the payoff for the original $m$-step strategy. This, in turn, implies that $m=1$ and the resulting problem reduces exactly to~\eqref{eq:optimal-arbitrage}.

\paragraph{Reported price.} The optimality conditions for problem~\eqref{eq:optimal-arbitrage} suggest a simple definition for the prices that should be reported by the CFMM, at some reserves $R'$: the CFMM should report the slope of the supporting hyperplane of $S(R)$ at the point $R'$, scaled appropriately by the num\'eraire. In other words, if $g \in \reals^n$ is a supporting hyperplane of $S(R)$ at $R'$, then the reported price should be $\lambda g$ for some $\lambda \ge 0$.

The scaling constant exists since the choice of num\'eraire is always left to the CFMM designer and does not change the optimality conditions. For example, if coin 1 is chosen as the num\'eraire, then the designer would choose $\lambda = 1/g_1$ and the CFMM would then report $g/g_1$ as the prices of all coins, such that the reported price of coin 1 is always $(g/g_1)_1 = 1$, as desired.

\begin{figure}
\centering
\begin{tikzpicture}
\begin{axis}[
axis lines=center, 
grid=major,
ymin=0, 
no marks,
xmin=0,
xmax=4,
xlabel=$R_2$,
ylabel=$R_1$,
x label style={at={(axis description cs:0.5,-0.1)},anchor=north},
y label style={at={(axis description cs:-0.1,.5)},rotate=90,anchor=south},
]
\addplot+[smooth, name path=A, domain=0.25:4, samples=100] {max(1/x, 1)};
\addplot+[draw=none, name path=B] {4};
\addplot+[blue, opacity=0.15] fill between[of=A and B, soft clip={domain=0.25:4}];
\addplot+[red, domain=0:2] {-1*(x-1) + 1} node[below, pos=0.7] {$g^2$}; 
\addplot+[red, domain=0:4] {1 - .1*(x-1)} node[below, pos=0.6] {${g^2}'$};
\addplot+[red, domain=0:1] {4*(1-x)} node[left, pos=0.4] {$g^1$};
\end{axis}

\tikzstyle{fontbf} = [thick, font=\fontsize{10}{0}\selectfont]
\draw (3.5, 3.5) node[fontbf] {$S(R)$};

\end{tikzpicture}
\caption{Reported prices for an example CFMM.}
\label{fig:price}
\end{figure}
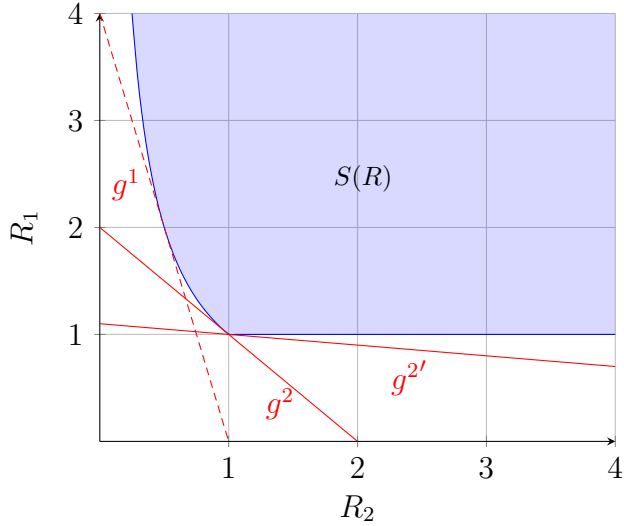

We also note that there may be many supporting hyperplanes of $S(R)$ at any one point (\cf, $g^2$ and ${g^2}'$ at reserves $(1, 1)$ in figure~\ref{fig:price}) in which case there are many no-arbitrage prices implied by these reserves and no unique price can be reported. For example, if there exist two supporting hyperplanes with slopes $g$ and $g'$, then any convex combination of $g$ and $g'$ is also a supporting hyperplane and therefore a valid price. For convenience, we will assume that all such prices are reported, though this need not be true in practice.

\paragraph{Discussion.} By reporting the price at the optimality conditions of the optimal arbitrage problem, we have essentially resolved two issues in one.

First, the definition of the reported prices implies that, given a reference market with prices equal to $c \in \reals_+^n$, an arbitrageur is always incentivized to make any one of the prices reported by the CFMM equal to $c$, since any other trade is strictly suboptimal. In other words, if the price of the CFMM is mismatched to that of a reference market, an agent is able to make what is essentially `free money' by only trading between these markets, which, in turn, would correct the price reported by the CFMM to be the same as the reference market price. 

Second, unlike the reported price, there is no guarantee that a marginal price at one specific reserve value is well-defined, since the definition of a marginal price requires the existence of arbitrarily small trades. (One simple example where a marginal price might not exist is to consider any CFMM that requires a fixed, nonzero amount of input before any trade is accepted.) On the other hand, if the CFMM is path independent, then the reported price and the marginal price will always match exactly. We show a proof in~\S\ref{app:marginal-price}.

\paragraph{Special cases.} The reported price can be easily computed in the common case where the trading set can be written in the form
of~\eqref{eq:sufficient-condition}, in which case the reachable set is constant after any optimal trade since every optimal trade is never in the dominated interior, which, in turn, implies that an optimal trade always leaves the reachable set unchanged by the definition of path independence. We can then use~\eqref{eq:sufficient-condition} to write the reachable sets as
\[
S(R) = S(R^0) = \{R' \in \reals^n \mid \psi(R') \ge \psi(R^0)\},
\]
for any $R \in S(R^0) \setminus \dint S(R^0)$, where $R^0$ are the reserves of the CFMM before the trade was completed. In this case, the first order optimality conditions of~\eqref{eq:optimal-arbitrage} would imply that
\[
c \in \lambda \partial \psi(R'^\star),
\]
where $\partial \psi(R'^\star)$ are the subgradients of the function $\psi$ at $R'^\star$, that are, by definition, supporting hyperplanes of the epigraph of $\psi$  (see~\cite[\S23]{rockafellar1970convex}). Here, $\lambda \ge 0$ acts as a scaling constant and depends on the choice of num\'eraire.

If $\psi$ is differentiable, then we would instead have (see Theorem~25.1 of~\cite{rockafellar1970convex}),
\[
c = \lambda \nabla \psi(R'^\star),
\]
so the reported prices of this CFMM at some reserves $R$ are simply proportional to $\nabla \psi(R)$. In general, when the trading set of a CFMM can be written in the form of~\eqref{eq:sufficient-condition}, we would expect the function $\psi$ to be differentiable since this case corresponds to the existence of exactly one possible price that can be
reported for some reserves $R$.
%
%Another important special case is when the trading set can be written as
%\[
%T(R) = \{(\Delta, \Lambda) \in \reals^n_+ \mid \psi(R, \Delta, \Lambda) \ge 0\},
%\]
%where $\psi$ is a quasiconcave function that is decreasing in $\Delta$ and increasing in $\Lambda$ with the null trade feasible, $(0,0) \in T(R)$. In this case, the optimality conditions for problem~\eqref{eq:optimal-arbitrage-trade-set} are that (see~\S\ref{app:optimality-conditions})
%\[
%(c, -c) \ge \lambda g, ~~ \text{for all} ~ g \in \partial \psi_R(\Delta^\star, \Lambda^\star),
%\]
%and the inequality holds at equality in the indices where $\Delta_i^\star \ne 0$ and $\Lambda_i^\star \ne 0$. Here, we have written $\psi_R$ for the function $\psi(R, \cdot, \cdot)$. The reported prices would then be proportional to any element in the set
%\[
%\{\lambda(g_\Delta, g_\Lambda) \in \partial \psi_R(\Delta^\star, \Lambda^\star) \mid g_\Delta = - g_\Lambda\},
%\]
%after appropriate scaling.

\paragraph{Practical examples.} In the case of constant mean markets (\eg, Balancer), given the reserves $R$, we can very easily
give the reported prices in the fee-less case. In this case, note that, from~\eqref{eq:balancer-cfmm},
\[
\psi(R) = \prod_{i=1}^n R_i^{w_i},
\]
so, after taking the gradient, we have:
\[
(\nabla\psi(R))_j = \frac{w_j}{R_j}\prod_{i=1}^n R_i^{w_i} = \frac{w_j}{R_j}\psi(R).
\]
Therefore, the price of some coin $j$ with respect to some second coin $k$ corresponds to choosing coin $k$ as the num\'eraire, \ie, choosing $\lambda = 1/(\nabla\psi(R))_k$, we get
\[
\lambda (\nabla\psi(R))_j = \frac{(\nabla\psi(R))_j}{(\nabla\psi(R))_k} = \frac{w_j/R_j}{w_k/R_k}.
\]
And, because $\gamma = 1$, the CFMM is path independent, the reported price is equal to the marginal price. Note that this result is equal to the original derivation of the marginal price given in the Balancer whitepaper~\cite[Eq.\ 7]{balancer}, though it is considerably more concise to derive using this formalism.

It is also not difficult to derive the prices that the Curve CFMM should report. From~\eqref{eq:curve-cfmm},
\[
\psi(R) = \alpha\ones^TR - \beta \left(\prod_{i=1}^n R_i\right)^{-1},
\]
so
\[
(\nabla \psi(R))_j = \alpha + \beta\left(R_j\prod_{i=1}^n R_i\right)^{-1},
\]
and, taking coin $k$ as the num\'eraire, we find the price of coin $j$ as,
\[
\frac{(\nabla \psi(R))_j}{(\nabla \psi(R))_k} = \frac{\alpha + \beta\left(R_j\prod_{i=1}^n R_i\right)^{-1}}{\alpha + \beta\left(R_k\prod_{i=1}^n R_i\right)^{-1}}.
\]

\subsection{Liquidity providers and returns}
An important aspect of CFMMs is the ability for liquidity providers, as described in~\S\ref{sec:cfmm-multi}, to add or remove
value from the reserves. Because liquidity providers own some explicit percentage of the total amount of coins in reserves,
we consider what the current value of the CFMM's total reserves are, under the same assumptions of~\S\ref{sec:arbitrage}:
that (a) there exists an external, infinitely liquid reference market and (b) there exists an arbitrageur who is solving
the arbitrage problem. The total value in the reserves is then proportional to the value of the position a liquidity provider
has, with respect to the CFMM (where the proportionality constant is simply the percentage of the reserves that a liquidity provider
is entitled to).

\paragraph{Value in reserves.} By definition, the total value of all assets held in reserves, $R$, by the CFMM is simply $c^TR$, where
$c_i > 0$ is the price of currency $i$ given by a reference market. To say what the total value of the reserves is, note that
problem~\eqref{eq:optimal-arbitrage} is equivalent to the following problem:
\begin{equation}\label{eq:reserve-value}
\begin{aligned}
	& \text{minimize} && c^TR'\\
	& \text{subject to} && R' \in S(R),
\end{aligned}
\end{equation}
with variable $R' \in \reals^n$, in the sense that an optimal $R'$ for~\eqref{eq:optimal-arbitrage} is optimal for~\eqref{eq:reserve-value} and vice versa. We can see this since the term $c^TR$ in the objective of~\eqref{eq:optimal-arbitrage} is a constant in the problem, and we've simply switched maximizing a function with minimizing its negative. Since this is true, then the optimal value of~\eqref{eq:reserve-value} is exactly the total value of the reserves.

Throughout the remainder of this section, we will let $p^\star_R(c)$ be the optimal value of problem~\eqref{eq:reserve-value} for a given cost vector $c$ and fixed reserves $R$. We also note that the function $p^\star_R(c)$ can be recognized as the negative of the support function of the set $S(R)$~\cite[\S13]{rockafellar1970convex}.

\paragraph{Path deficiency.} An important consequence of path deficiency with respect to the total reserve value is that, in general, the total value of the reserves never decreases for a fixed cost vector. In particular, if $R$ are the reserves at some (initial point) and $R'$ are the reserves at
some future point (after any number of trades have been performed), then, by definition of path deficiency, we have $S(R') \subseteq S(R)$. This immediately implies that $p^\star_{R'}(c) \ge p^\star_R(c)$, since problem~\eqref{eq:reserve-value} with reserves $R$ contains all of the same feasible points of problem~\eqref{eq:reserve-value} with reserves $R'$, so its optimal value can be no higher. It is not difficult to prove that strict path deficiency implies that this value is, instead, strictly increasing---the idea follows
from the fact that all optimal points of~\eqref{eq:reserve-value} with reserves $R$ will always lie in $S(R) \setminus S(R')$ and that $S(R')$ is a closed set.

\paragraph{The dual function.}
If we can write the reachable reserve set as,
\[
S(R) = \{R' \in \reals^n \mid \psi(R') \ge \psi(R)\},
\]
for some quasiconcave, increasing function $\psi$, then the Lagrangian~\cite[\S5.1.1]{cvxbook} of problem~\eqref{eq:reserve-value} is
\[
\mathcal L(R', \lambda) = c^TR' - \lambda (\psi(R') - \psi(R)),
\]
which lets us write the dual function $g(\lambda) = \inf_{R'} \mathcal L(R', \lambda)$ for $\lambda \ge 0$. This function, written in terms of $\psi$ is
\[
	g(\lambda) = \lambda \psi(R) + \inf_{R'}\left( c^TR' - \lambda \psi(R')\right) \\
	= \lambda \psi(R) - \sup_{R'}\left( (-c)^TR' - \lambda(-\psi)(R')\right).
\]
If $\lambda > 0$, we have that the last term is equal to
\[
\sup_{R'}\left( (-c)^TR' - \lambda(-\psi)(R')\right) \\
= \lambda \sup_{R'}\left( (-c/\lambda)^TR' - (-\psi)(R')\right) = \lambda \left(-\psi\right)^* \left(-\frac{c}{\lambda}\right),
\]
where $(-\psi)^*$ is the Fenchel conjugate~\cite[\S3.3]{cvxbook} of $-\psi$ and is known for a very large number of convex functions. We also note that this operation is just the perspective transform of $(-\psi)^*$ (sometimes called epi-multiplication~\cite[\S5]{rockafellar1970convex}) and is well defined even when $\lambda = 0$, though we will continue to write the term as is for simplicity. Combining the resulting statements, we get
\[
g(\lambda) = \lambda \psi(R) - \lambda \left(-\psi\right)^* \left(-\frac{c}{\lambda}\right),
\]
for $\lambda \ge 0$.

By weak duality~\cite[\S5.2.2]{cvxbook}, we know
that $\sup_{\lambda \ge 0} g(\lambda) \le p^\star_R(c)$, so picking any $\lambda \ge 0$ will suffice to give a lower bound on the total value of the reserves. Additionally, since $\psi$ is strictly monotonic, then strong duality holds~\cite[\S5.2.3]{cvxbook} and we have that
\begin{equation}\label{eq:total-reserve-values}
p^\star_R(c) = \sup_{\lambda \ge 0} g(\lambda) = \sup_{\lambda \ge 0} \left(\lambda \psi(R) - \lambda \left(-\psi\right)^* \left(-\frac{c}{\lambda}\right)\right),
\end{equation}
instead. This construction gives us a way of computing total reserve values by solving a single-variable convex optimization problem (since $g$ is always a convex function~\cite[\S5.1.2]{cvxbook}) and, in some important cases, gives closed form expressions for the total reserve value.

\paragraph{Constant mean reserve values.} Using the constant mean trade function with $\gamma = 1$, given in~\eqref{eq:balancer-cfmm},
we can write
\[
\psi(R) = \prod_{i=1}^n R_i^{w_i}.
\]
We will write, for convenience, $k = \psi(R^0)$ at the initial reserves $R^0$. Note that, since this CFMM is path independent, the
reachable reserve set does not change and we may assume that the reachable reserve set is simply $S(R) = \{R'\in\reals^n\mid \psi(R') \ge k\}$ for any reachable reserves $R$.

From~\cite[prob.\ 3.36]{cvxbook}, we have that, for any cost vector $c \in \reals_+^n$:
\[
(-\psi)^*(-c) = \begin{cases}
0 & ~\prod_{i=1}^n c_i^{w_i} \ge 1/n\\
\infty & \text{otherwise}.
\end{cases}
\]
As before, the current value of the reserves is given by~\eqref{eq:total-reserve-values}:
\[
\sup_{\lambda \ge 0} \left(\lambda k - \lambda \left(-\psi\right)^* \left(-\frac{c}{\lambda}\right)\right).
\]
Now, note that
\[
\lambda k - \lambda \left(-\psi\right)^* \left(-\frac{c}{\lambda}\right) = \begin{cases}
\lambda k & \left(\prod_i c_i^{w_i}\right) /\lambda \ge 1/n\\
-\infty & \text{otherwise},
\end{cases}
\]
which we can easily maximize by choosing the largest possible $\lambda$, \ie, $\lambda = n\prod_i c_i^{w_i}$, since $k > 0$, yielding
\begin{equation}\label{eq:balancer-returns}
\sup_{\lambda \ge 0} \left(\lambda k - \lambda \left(-\psi\right)^* \left(-\frac{c}{\lambda}\right)\right) = nk\prod_{i=1}^n c_i^{w_i}.
\end{equation}

Using the special case of $n=2$ and $w_1 = w_2 = 1/2$, we can recover the total reserve value of Uniswap with zero fees, derived via a different method in~\cite[\S2.3]{angeris2019analysis}. To see this, set $c_1 = 1$ (\ie, coin 1 is the num\'eraire) such that $c_2$ is the market price of coin 2 with respect to coin 1. Then, we can write the total value in reserves as,
\[
p^\star = 2k\sqrt{c_2},
\]
where the square root difference between the expression in~\cite[\S2.3]{angeris2019analysis} and this one comes from the fact that $k$ here is the \emph{square root} of the product constant, \ie, $k = \sqrt{R_1R_2}$. (See~\cite[\S3]{angeris2019analysis}.)

\paragraph{Lower bounds for Curve.} We suspect that there is no analytical solution to the total reserve value for the Curve CFMM, given
in~\eqref{eq:curve-cfmm}. We will, instead, derive some lower bounds to the total reserve value by appropriately choosing $\lambda \ge 0$, where the function we consider is
\[
\psi(R) = \alpha\ones^TR - \beta \left(\prod_{i=1}^n R_i\right)^{-1},
\]
which means the CFMM's reachable set is $S(R) = \{R \in \reals^n \mid \psi(R) \ge k\}$,
where $k = \psi(R^0)$ and this CFMM is path independent.

First, for $\lambda \ge 0$,
\[
\lambda (-\psi)^*\left(-\frac{c}{\lambda}\right) = -(n+1)(\lambda\beta)^{1/(n+1)}\prod_{i=1}^n (c_i - \lambda\alpha)^{1/(n+1)},
\]
if $c_i - \lambda\alpha \ge 0$ for each $i=1, \dots, n$ and $+\infty$, otherwise (see~\S\ref{app:reciprocal-conjugate}). The dual function, $g: \reals \to \reals$, is then
\[
g(\lambda) = \lambda k + (n+1)(\lambda\beta)^{1/(n+1)}\prod_{i=1}^n (c_i - \lambda\alpha)^{1/(n+1)},
\]
if $c_i - \lambda\alpha \ge 0$ for each $i=1, \dots, n$ and $-\infty$, otherwise. Some simple lower bounds come from considering
$\lambda$ as large as possible, which is somewhat tight if $k$ is very large relative to $(\beta\prod_{i=1}^nc_i)^{1/(n+1)}$.
We can do this by choosing $\lambda = \min_i c_i/\alpha$, giving the following lower bound for the total reserve value:
\[
p^\star_R(c) \ge \left(\frac{k}{\alpha}\right)\min_ic_i.
\]
Though we suspect that, in general, the exact value of $p^\star_R(c)$ cannot be given in closed form, we note that the resulting problem of optimizing $g$ is a one-parameter convex optimization problem that is, in practice, easy to solve numerically.

\section{Extensions and future work}
There are several important applications and essentially immediate extensions of the current conditions and definitions that are interesting to study in their own right. We discuss some basic examples here.

\subsection{Trading fees}\label{sec:trading-fees}
We can easily introduce trading fees to any given CFMM. A simple but effective approach is to introduce fees on the input trade. Given a trading function $\phi$ for a CFMM, we can then write a new trading function $\phi_f$ with some fee constant $0 < \gamma \le 1$ defined as
\[
\phi_f(R, \Delta, \Lambda) = \phi(R, \gamma \Delta, \Lambda),
\]
for all reserves $R$, inputs $\Delta$, and outputs $\Lambda$ where $(1-\gamma)$ is the percentage fee required. Equivalently, we can write this in terms of a new trading set $T_f(R)$ as
\[
T_f(R) = \{(\Delta, \Lambda) \mid (\gamma\Delta, \Lambda) \in T(R)\},
\]
for each reserve $R$.

In this case, the trader is required to put in $1/\gamma$ more of input $\Delta$ for a trade to be feasible. Additionally, this will turn path independent CFMMs into path deficient ones (strictly path deficient, if $\gamma < 1$). Because the resulting CFMM is path deficient, this method also has the nice property that the total reserve values are always bounded from below by the solution to~\eqref{eq:reserve-value}.

There are several more possible methods, some of which may include variable input and/or output fees which vary in such a way as to keep other desirable properties of the CFMMs on a case-to-case basis. We suspect that there are many approaches for charging trading fees, each with their own useful properties, but leave the possibility of finding a suitable class of these trading fees that is good to study for future work.

\subsection{Comparison to scoring rules}\label{sec:lmsr-comparison}
While it is tempting to ask about potential comparisons or equivalences to classic algorithmic game theory automated market makers and scoring rules such as Hanson's LMSR or the more general constant utility market makers for prediction markets provided in~\cite{othmanAutomatedMarketMakers2012}, we note that this is likely not possible using only the no-arbitrage framework used in this paper. A simple thought experiment shows why this might be the case.

Given an infinitely liquid reference market with a fixed price $p_1$ from time $[0, T)$, where $T>0$, and price $p_2 > p_1$ at time $T$ (known by all agents), then the reported price of a prediction market which seeks to predict the price of the asset at time $T$ and the reported price of a given CFMM will always diverge by $p_2 - p_1$. Rational agents will always be incentivized to correctly report the (known) future price $p_2$ for the prediction market, while arbitrageurs will always make positive payoff from any CFMM which diverges from the current market price $p_1$ at all times $[0, T)$, by setting the reported price to be $p_1$. Sending $p_2 - p_1 \to \infty$ then shows that these two AMMs can diverge by any desired amount.

The idea here is that any framework which can compare the two will require assumptions about the market price dynamics, \ie, what the current market price might say about the future market price, which we do not assume at any point in this presentation. We leave this potentially very interesting research avenue of finding a suitable framework for comparison for future work.

\subsection{Optimization over possible CFMMs}
Note that the given conditions define a family of CFMMs which are likely to be useful in practice. This implies that, for any performance metric (\eg, average total reserve value for a given market model) that a market maker designer wishes to optimize, one could find an (approximately) optimal CFMM to accomplish this task. The problem is likely to be computationally difficult to solve exactly in most important cases, but we suspect that many commonly-used heuristics will likely find good results. Though this approach is unlikely to be feasible except when $n$ is small, we imagine that the very useful case of $n=2$ can be quickly optimized on modern hardware for many useful performance metrics. 

Additionally, if the trading function $\phi$ is parametrized by a small number of parameters (for example, the parameters $\alpha$, $\beta$ in~\eqref{eq:curve-cfmm}), it is possible to at least approximately optimize these parameters to maximize or minimize some desired objective function of the trading sets or the reachable sets.

\subsection{Time-dependent CFMMs and other generalizations}
Note that the conditions and definitions above can be very easily extended to the cases where the trading function $\phi$ depends on exogenous variables such as time. In particular, we may assume that arbitrage happens instantaneously or nearly instantaneously. This, in turn, would imply that some (but not all) of our analysis on general CFMMs holds even in this scenario.

There are still several interesting questions to answer in this case. One such question is: what is a natural generalization of
the reachable set when the trading function (or, equivalently, the reachable set) is also time-dependent? A simple (but likely woefully incomplete) answer is that, if the reachable set depends on time, say $S_t(R)$, where $t \ge 0$ is a time variable, we additionally have $S_{t'}(R) \subseteq S_t(R)$ for all $R$ and $t' \ge t$. This retains some of the given properties, such as the lower bounds on the total reserve values given in~\eqref{eq:total-reserve-values}, but is likely to be too strong of a condition to be useful in practice.

Another natural question is, are there good restrictions on how liquidity providers should add liquidity to reserves? One could imagine that, in some scenarios, allowing agents to add coins to reserves in an arbitrary way could lead to large losses for liquidity providers. An even more fundamental question, which we do not cover at all is: can liquidity provision easily be included in a similar framework? We suspect so, but even this is not clear at the moment and is likely to be a good avenue for future exploration.

\section{Conclusion}
The increase in usage and participation in automated market makers has led to a vast set of new scoring rules and pricing mechanisms.
Analyzing these mechanisms, which range from LMSR style market makers and CFMMs to scoring rules for rates~\cite{chitra2019competitive}, from the perspective of optimization provides insight into why certain mechanisms are more popular than others and work well in practice.
In particular, we show that CFMMs provide an easy optimization problem for arbitrageurs to synchronize off-chain and on-chain pricing data, along with several useful conditions that often hold in practice, which imply that CFMMs are likely to be very well behaved.
This generalization encompasses all live CFMMs \cite{uniswap, balancer, bancor, egorovStableSwapEfficientMechanism} and provides guidance on how one can design CFMMs that are better for certain asset types and volatilities.
%
%Further work on CFMMs is necessary to connect how to choose the optimal scoring rule given the volatilities of the external price feeds.
%As the StableSwap/Curve example demonstrates, adjusting the curvature of the rule can lead to improved performance and participation for different assets.
%Finally, we leave the study of how to estimate the cost of corruption~\cite{angeris2019analysis, lambur_lu_cai_2019}, which can be thought of as a relaxation of the price of anarchy~\cite{roughgarden2005selfish}, for future work. 

\section*{Acknowledgements}
%\begin{acks}
We would like to thank John Morrow and Tim Roughgarden for feedback on this paper and Alex Evans for feedback and pointing out that the current CFMM analysis extends to the case where the trading function is time dependent.
%\end{acks}

%\bibliographystyle{ACM-Reference-Format}
\appendix
\section{Miscellaneous Proofs}
\subsection{An equivalent monotonic trade function}
\label{app:monotonicity}
Given some trade function $\phi$, we will write an equivalent CFMM with trading function $\phi'(R, \cdot, \cdot)$ that is monotonically nonincreasing in its first argument and monotonically nondecreasing in its second, for each possible reserves $R$.

First, define the squared distance-to-set function for some set $U \subseteq \reals^n$ and point $x \in \reals^n$ as
\[
d(x, U) = \inf_{y \in U} \|x - y\|_2^2.
\]
Then the following function
\[
\phi'(R, \Delta, \Lambda) = d((\Delta, \Lambda), T(R)) + d(0, T(R)),
\]
suffices. In this case, $\phi'(R, \cdot, \cdot)$ measures the squared distance of a given trade to the trade set, with some offset given by $d(0, T(R))$.

\paragraph{Equivalence.} First, if $\phi'(R, \Delta, \Lambda) = \phi'(R, 0, 0)$ then we have that $d((\Delta, \Lambda), T(R)) = 0$ so $(\Delta, \Lambda) \in T(R)$, since $T(R)$ is a closed set. Conversely, if $\phi'(R, \Delta, \Lambda) \ne \phi'(R, 0, 0)$ then $\phi'(R, \Delta, \Lambda) > \phi'(R, 0, 0)$ since the squared distance function is nonnegative, which means $d((\Delta, \Lambda), T(R)) > 0$ and, in turn, that $(\Delta, \Lambda) \not\in T(R)$. This implies that the function $\phi'$ has trading set $T(R)$ and is therefore equivalent to $\phi$. (As a second useful note, if the set $T(R)$ is convex, then $\phi'(R, \cdot, \cdot)$ is also convex in its arguments~\cite[\S3.2.5]{cvxbook}.)

\paragraph{Monotonicity.} Note that, since monotonicity is defined elementwise, it suffices to prove monotonicity for general sets $Q\subseteq \reals \times \reals^n$ of the form
\[
Q = \{(t, q) \mid (t', q) \in W ~\text{for some}~ t' \le t\},
\]
for some nonempty closed set $W \subseteq \reals \times \reals^n$. (Note that the set $Q$ is closed if the set $W$ is.)

We will show that the squared distance-to-set function, $d(\cdot, Q)$, is decreasing with respect to the first element of its first argument. Consider some pair $(t, q) \in \reals\times \reals^n$ and let $(t^\star, q^\star)\in \reals\times \reals^n$ minimize $d(t, q)$, which exists since the set is closed. Now, for any $t' \in \reals$ with $t' \ge t$, either $t' \ge t^\star$, in which case 
\[
d(t', q) \le \|q - q^\star\|_2^2 \le d(t, q)
\]
since $(t', q^\star) \in Q$ by definition of $Q$, or $t' < t^\star$, which implies that $t \le t' < t^\star$ so clearly,
\[
d(t', q) \le (t' - t^\star)^2 + \|q - q^\star\|_2^2 < (t - t^\star)^2 + \|q - q^\star\|_2^2 =  d(t, q),
\]
which shows that $d(t', q) \le d(t, q)$ for every $t' \ge t$.

The complete proof then follows from the fact that we can apply this proof elementwise to the function $\phi'(R, \cdot, \cdot)$.

\paragraph{Discussion.}
In general, the fact that we can always write a monotonic trade function for any trade set implies that problem~\eqref{eq:optimal-arbitrage-trade-set}
and the following `relaxed' problem are equivalent (for fixed reserves $R$),
\begin{equation}\label{eq:relaxed-formulation}
\begin{aligned}
	& \text{minimize} && c^T(\Lambda - \Delta)\\
	& \text{subject to} && \phi'(R, \Delta, \Lambda) \le 0.
\end{aligned}
\end{equation}
Here, the variables are $\Lambda, \Delta \in \reals^n$ and the problem data are $R$ and $c$.

To see this, first note that, any trade $(\Delta, \Lambda)$ that is feasible for problem~\eqref{eq:optimal-arbitrage-trade-set} is clearly feasible for this problem
since, by definition of $\phi'$, $\phi'(R, \Delta, \Lambda) = 0$, so the optimal objective value is no larger than that of~\eqref{eq:optimal-arbitrage-trade-set}.
Additionally, no optimal point $(\Delta^\star, \Lambda^\star)$ for problem~\eqref{eq:relaxed-formulation} will ever satisfy $\phi'(\Delta^\star, \Lambda^\star) < 0$. We can see this since the function $\phi'$ is continuous, which implies that the set of all $(\Delta, \Lambda)$ satisfying $\phi'(\Delta, \Lambda) < 0$ is an open set. This, in turn, implies that, for any trade $(\Delta, \Lambda)$ satisfying $\phi'(\Delta, \Lambda) < 0$ there always exists a trade $(\Delta', \Lambda')$ satisfying $\Delta' < \Delta$ and $\Lambda' > \Lambda$ (since there always exists such a pair $(\Delta', \Lambda')$ in some neighborhood around $(\Delta, \Lambda)$, for a nonempty open set) which clearly has strictly lower objective value, so $(\Delta, \Lambda)$ cannot be optimal for~\eqref{eq:relaxed-formulation}. So, $\phi'(R, \Delta^\star, \Lambda^\star) = 0$ and, therefore, that we also have $(\Delta^\star, \Lambda^\star) \in T(R)$, so any optimal point for~\eqref{eq:relaxed-formulation} is also feasible for~\eqref{eq:optimal-arbitrage-trade-set}, implying that both problems have the same optimal value.

Because $\phi'(R, \cdot, \cdot)$ is convex if $T(R)$ is, then problem~\eqref{eq:relaxed-formulation} is additionally a convex optimization problem whenever
$T(R)$ is convex.

\subsection{Marginal price for path independent CFMMs}\label{app:marginal-price}
First, we define the price of performing a trade as the minimal cost of receiving a desired output of $\Lambda$, where the
value of coin $i$ is given by $c_i > 0$:
\begin{equation}\label{eq:marginal-price}
p(R, c, \Lambda) = \inf \{c^T\Delta \mid R + \Delta - \Lambda \in S(R)\}.
\end{equation}

Now, under the scenario given in problem~\eqref{eq:optimal-arbitrage}, we have that the reported price, $c$ is a supporting
hyperplane of $S(R)$ at the point $R' = R + \Delta^\star - \Lambda^\star$ where $(\Delta^\star, \Lambda^\star)$ is the optimal arbitrage trade, and that $S(R') = S(R)$, since the CFMM is path independent. We can then define the marginal price of a desired output $\Lambda$, after the optimal trade is completed, as the limit:
\[
\lim_{\eps \downto 0} \frac{p(R', c, \eps\Lambda)}{\eps}.
\]
By definition, since $c$ is a supporting hyperplane of $S(R)$ at $R'$, we have that
\[
c^T(R'' - R') \ge 0, ~ \text{for all} ~~ R'' \in S(R),
\]
which implies that
\[
c^T(\Delta - \Lambda) \ge 0,
\]
for all $\Delta, \Lambda$ satisfying $R' + \Delta - \Lambda \in S(R) = S(R')$, which
can easily be rewritten as $c^T \Delta \ge c^T\Lambda$. Using this inequality with the definition of the marginal price~\eqref{eq:marginal-price}, we find:
\[
\frac{p(R', c, \eps\Lambda)}{\eps} \ge \frac{1}{\eps} (\eps c^T\Lambda) = c^T\Lambda.
\]
but the inequality is, in fact, an equality since $\Delta = \eps\Lambda$ is a feasible point
for~\eqref{eq:marginal-price} as $R' \in S(R) = S(R')$. Since the definition of the marginal price of coin $i$ is the average
price of buying an infinitesimally small quantity of coin $i$; \ie,
\[
\lim_{\eps \downto 0} \frac{p(R', c, \eps e_i)}{\eps} = c_i,
\]
where $e_i$ is the $i$th unit vector, then the price reported matches the marginal price at reserves $R'$, as required.

\subsection{Conjugate of reciprocal product}\label{app:reciprocal-conjugate}
We show here that the convex function given by
\[
f(x) = \left(\prod_{i=1}^n x_i\right)^{-1},
\]
has a Fenchel conjugate given by
\[
f^*(y) = \begin{cases}
	-(n+1)\left(\prod_{i=1}^n (-y_i)\right)^{1/(n+1)} & y \le 0\\
	+\infty & \text{otherwise}.
\end{cases}
\]

The proof is mostly mechanical. The definition of the Fenchel conjugate of the function $f$ is given by
\[
f^*(y) = \sup_x \,(y^Tx - f(x)),
\]
where $f(x)$ is extended in the natural way (\ie, $f(x) = +\infty$ if $x_i \le 0$ for any $i$). First, we show that if $y_i > 0$ for some $i$, then $f^*(y) = +\infty$. To do this, note that we can set $x_i = t$ and $x_j = 1$ for every $j$ with $j \ne i$. Then sending $t \to \infty$, we get that $y^Tx \to \infty$, while $f(x) \to 0$, implying the result.

Now we consider the case where $y \le 0$. First, note that, in this case
\[
y^Tx - f(x) \le 0,
\]
for any $x$, since $x > 0$ and therefore $f(x) \ge 0$ (otherwise, if $x \not > 0$, we have $f(x) = +\infty$ by definition, so the claim is always satisfied). Now, if there exists some $i$ with $y_i = 0$, then we can achieve this bound by setting $x_i = t^n$ and $x_j = 1/t$ for $j \ne i$. Sending $t \to \infty$ yields the result since $y^Tx \to 0$ and $f(x) \to 0$.

The remaining case is when $y < 0$. Here, we can write the first order optimality conditions over $x$ (which are sufficient and necessary by convexity and differentiability) after some simplifications:
\[
x_i^\star y_i = -\left(\prod_{j=1}^n x_j^\star\right)^{-1}, \quad i=1, \dots, n,
\]
or, equivalently, that
\[
x_i^\star\prod_{j=1}^n x_j^\star = \frac{1}{-y_i}, \quad i=1, \dots, n.
\]
This implies that $x$ is collinear with the negative reciprocal of $y$, \ie, that $x_i^\star = \lambda / (-y_i)$, and, since $x^\star \ge 0$, we must have $\lambda \ge 0$. This implies
\[
\lambda^{n+1} = \prod_{i=1}^n(-y_i),
\]
and therefore,
\[
x^\star_i = \frac{1}{-y_i}\left(\prod_{i=1}^n(-y_i)\right)^{\frac{1}{n+1}},
\]
which implies that
\[
y^Tx^\star - f(x^\star) = -(n+1)\left(\prod_{i=1}^n (-y_i)\right)^{\frac{1}{n+1}},
\]
completing the proof.

\bibliographystyle{unsrt}
\bibliography{bib}

\end{document}

%% file: defs.tex
\newcommand{\ones}{\mathbf 1}
\newcommand{\reals}{{\mbox{\bf R}}}

\newcommand{\relint}{\mathop{\bf relint}}

\newcommand{\cf}{{\it cf.}}
\newcommand{\eg}{{\it e.g.}}
\newcommand{\ie}{{\it i.e.}}

\newcommand{\BEAS}{\begin{eqnarray*}}
\newcommand{\EEAS}{\end{eqnarray*}}
\newcommand{\BEA}{\begin{eqnarray}}
\newcommand{\EEA}{\end{eqnarray}}
\newcommand{\BEQ}{\begin{equation}}
\newcommand{\EEQ}{\end{equation}}
\newcommand{\BIT}{\begin{itemize}}
\newcommand{\EIT}{\end{itemize}}